\documentclass[11pt]{article}

\usepackage[english]{babel} 
\usepackage[utf8]{inputenc}
\usepackage[T1]{fontenc}

\usepackage[gnuplot36beta]{egplot}
\usepackage{amsmath}
\usepackage{amssymb} 
\usepackage{amsfonts}
\usepackage{mathtools} \usepackage{multicol}
\usepackage{mhchem}
\usepackage{float} 
\usepackage{bm}
\usepackage{mathrsfs}
\DeclareMathOperator{\sech}{sech}

\newcommand{\diff}{\,\mathrm{d}}
\newcommand{\reali}{\mathbb{R}}
\newcommand{\Def}{\stackrel{\mathrm{def}}{=}}

\title{A replacement of the Lorentz law for the shape of the spectral
lines in the infrared region}
\author{A. Carati \and A.M. Maiocchi}
\date{\today}

\begin{document}

\maketitle

\begin{egpfile}
\egpfigprelude{ 
  set term post enhanced eps dashed  size 7.9cm, 7.9cm  font "Garamond,22";
  set format y  "\%\# 1.e";
  set ylabel offset 2,0;
  set pointsize 1;
  set samples 10000 ;
  a = 0.00175  ;
  b = 0.082    ;
  c = 2.1e-07  ;
  f(x) = a*b*(b*b-x*x)/( (b*b-x*x)**2 + 4.*c*x*x)  ;
  chiLi=0.96;
  chiNa=1.36;
  chiMg=0.89;
}

\begin{abstract}
We propose a new phenomenological law for the shape of the
spectral lines in the infrared, which accounts for the exponential
decay of the extinction coefficient in the
high frequency region, observed in many spectra. We apply this law to the
measured infrared spectra of \ce{LiF}, \ce{NaCl} and \ce{MgF2},
finding a good agreement, over a wide range of frequencies.
\end{abstract}

\section{Introduction}

At a phenomenological level, the experimental data for the complex
susceptibility $\hat \chi(\omega)$ are fitted by taking for each line
a contribution of the form
\begin{equation}\label{eq:1}
\hat \chi(\omega) = \frac {\omega_0A}{\omega^2-\omega_0^2 + 2i\gamma\omega}  \ ,
\end{equation}
where $\omega_0$ is the line frequency, $\gamma$ is related to its
width and $A$ to its intensity. This formula was originally obtained
by thinking of each line as corresponding to a ``physical''
microscopic dipole oscillating with frequency $\omega_0$ and with a
damping characterized by the  constant 
$\gamma$. In the literature this is often referred to as the  ``Lorentz
model''. However, some  difficulties arise in connection with
the imaginary part of the complex susceptibility, which, in the
transparency region of  dielectrics,  dictates the behavior of the
extinction coefficient $\kappa$.\footnote{As is well known,  the
  susceptibility $\chi$ is 
  related to the extinction coefficient $\kappa$ and the 
  refractive index $n$ by $4\pi\,\mathrm{Re}\, \hat \chi =n^2 - \kappa^2 -1$ and
  $4\pi\,\mathrm{Im}\, \hat \chi =2n\kappa$. So, in the region
  in which the dielectric is transparent, i.e., where $n$ is
  approximately constant, the behavior of $\mathrm{Im}\, \hat \chi$
  determines the behavior of 
  $\kappa$.\label{foot:1}}  
In fact, it is known since the seventies that for dielectrics, in the region
of high transparency, the Lorentz formula  
(\ref{eq:1}) provides  for the extinction coefficient not only a too
large value (by orders of magnitude), but also a
qualitatively incorrect behavior. 

Indeed, relation (\ref{eq:1}) gives, for the imaginary part,
\begin{equation}\label{eq:1im}
\mathrm{Im}\, \hat \chi(\omega) = \frac {2 \gamma \omega_0 \omega A
}{(\omega^2-\omega_0^2)^2 + 4\gamma^2\omega^2} \ ,
\end{equation}
which reduces to the well known Lorentz formula
\begin{equation}\label{eq:Lorentz}
\mathrm{Im}\, \hat \chi(\omega) \simeq \frac { \gamma A/2
}{(\omega-\omega_0)^2 + \gamma^2} \ ,
\end{equation}
for frequencies near the absorption peak, $\omega\simeq\omega_0$.
So, formula (\ref{eq:1im}) predicts that the extinction coefficient
decreases as $\omega^{-3}$ for large $\omega$, whereas, in the
transparency region, i.e., for
$\omega\neq\omega_0$,  the measured values of the
extinction coefficient (see \cite{rupprecht,deutsch}) exhibit a
decay which is exponential rather than as an inverse power of $\omega$. 
In the literature (see \cite{sparks1,sparks2}) one finds involved ab
initio computations which reproduce quite well the experimental
findings, but a simple general reason for the observed behavior is
lacking.

In this paper we propose an explanation of the observed exponential
decay of the extinction coefficient, as due to the fact that the time
auto--correlation of polarization should be an analytic function of
time. We also propose a simple phenomenological formula which should be 
substituted for the Lorentz one, in order to describe the exponential
decay of the experimental data. Such a formula involves the
asymptotic (in time) behavior of the time auto--correlation of
polarization, as described below. 

In Section~\ref{sez:secante} the theoretical argument is presented and
the corresponding proposed formula is given. In
Section~\ref{sez:fit}  a quantitative check of the
proposed formula is performed, by fitting the experimental data for three
dielectrics (\ce{LiF}, \ce{NaCl},  \ce{MgF2}) over a very large interval of
frequencies in the infrared region. In the last Section~\ref{sez:com}
some comments are added, in particular concerning the relaxation of
the correlations of polarization.

\section{The susceptibility according to linear response theory}\label{sez:secante}

In modern terms, the Lorentz law (\ref{eq:1}) can be justified through linear
response theory as follows. The susceptibility is
nothing but the Fourier transform of the time correlation of polarization.
In formul\ae, one has\footnote{We are here
  considering isotropic systems, otherwise a
susceptibility tensor should be considered, and the formul\ae\ should be
 changed accordingly.} 
\begin{equation}\label{eq:def_chi}
\hat \chi(\omega) = \frac{V}{4\pi k_B T}\int_0^{+\infty} e^{-i\omega t} \langle P(0)\dot
P(t)\rangle \diff t ,
\end{equation}
where $P(t)$ is the system polarization and the brackets denote a suitable
average (for example the canonical one).
Now, the classical formula (\ref{eq:1}) is
obtained simply by integration if one supposes that the correlation
decays exponentially for $t\to 
+\infty$ as a damped sinusoid, i.e., as proportional to
$$
 \langle P(0)\dot P(t)\rangle \propto - \sin \omega_0 t\, e^{-\gamma t} \ .
$$
However, if 
at first sight such expression for the correlation can appear
physically sound,
one has to recall that the correlation $\langle P(0)\dot P(t)\rangle$ is
an odd function of time (because it is the derivative of the 
correlation $\langle P(0) P(t)\rangle$, which has to be even). Thus
the correct expression should rather be 
\begin{equation}\label{eq:corr_nonanalitica}
 \langle P(0)\dot P(t)\rangle \propto - \sin \omega_0 t \,e^{-\gamma |t|} \ ,
\end{equation}
which is \textbf{not} an analytic function of time, due to the
presence of $|t|$ in the argument of the exponential. 

This rather abstract remark immediately implies that the imaginary
part of susceptibility, and thus the extinction coefficient, cannot decay 
exponentially if the correlation has the form
(\ref{eq:corr_nonanalitica}). In fact, from (\ref{eq:def_chi}) one has 
$$
\mathrm{Im}\, \hat \chi(\omega) =  -\frac12\, \frac{V}{4\pi k_B T}
\int_{\reali} \sin \omega t \langle P(0)\dot 
P(t)\rangle \diff t \ ,
$$
which shows that the extinction coefficient decays exponentially if
and only if  $ \langle P(0)\dot P(t)\rangle $ is analytic as a
function of time.

On the other hand,  there is no reason for the appearance of a
singularity at $t=0$  in the expression of the correlation, and it
seems instead natural to suppose that $ \langle
P(0)\dot P(t)\rangle $ should be taken as an analytic function. So the problem 
is now reduced to finding a simple analytic function which is odd and
decays exponentially. One should however take into account the fact
that $\langle P(0)\dot P(t)\rangle$ is a correlation, in 
particular  the 
derivative of the time auto--correlation $\langle P(0) P(t)\rangle$, which is a
positive--definite function (in the sense of Bochner, see
\cite{boch}, p. 17).
This apparently abstract mathematical requirement in
particular implies that the extinction coefficient be  positive at all
frequencies,\footnote{In fact, since $\langle P(0) P(t)\rangle$ is
  positive--definite, one has 
  $$
  \langle P(0) P(t)\rangle = \int_{\mathbb R} d \omega \alpha(\omega)
  \cos(\omega t) \ ,
  $$
  with  $ \alpha(\omega)\ge 0$, $\forall \omega   \in \mathbb R$, in
  virtue of Bochner theorem. Then,
  $$
  \langle P(0) \dot P(t)\rangle = -\int_{\mathbb R} d \omega\, \omega
  \alpha(\omega) \sin(\omega t)\  ,
  $$
  so that  $\mathrm{Im} \hat
  \chi(\omega)=A \,\omega \alpha(\omega)$ with a suitable constant $A>0$, 
  and so $\mathrm{Im} \hat   \chi$ is positive for $\omega\ge 0$. 
} a very sound physical constraint indeed. 
\begin{figure}[!t]
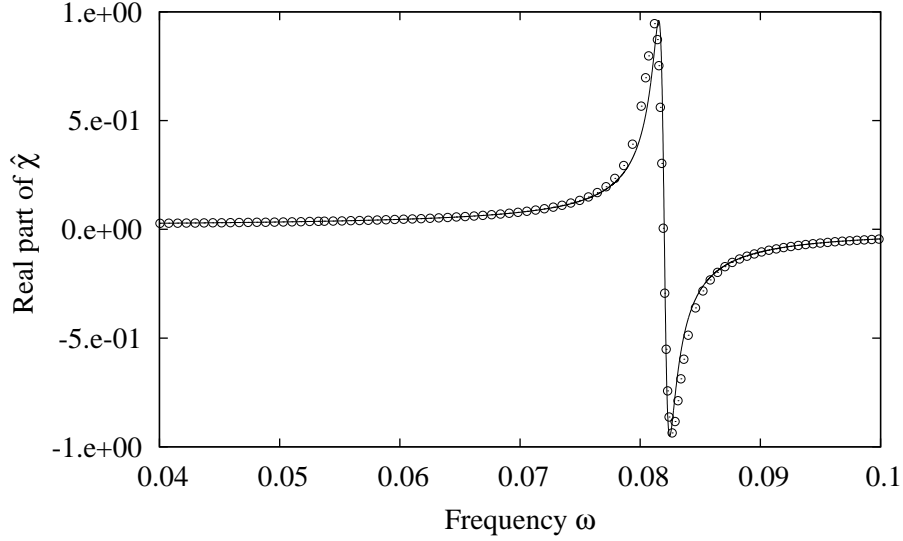

  \begin{center}
    \begin{egp}
      set term post enhanced eps dashed  size 12.5cm, 7.5cm  font "Times-Roman,22"
      set ylabel "Real part of ~{/Symbol c}{.1\303} "
      set xlabel "Frequency {/Symbol w}"
      set ytics -1.,.5,1.
      plot [0.04:0.1][-1.:1.] "./lorenziana_es_dati.txt" u 1:2 not w p pt 6,  f(x) not w l lt 1 
    \end{egp}  
  \end{center}
\caption{\label{fig:2} Real part of the Laplace transform of the
  function at the r.h.s. of (\ref{eq:funzione_prova})  
  computed numerically (circles), together with the real part of 
the Lorentz expression for $\hat \chi$ as given in (\ref{eq:1n}),
solid line. Here the relevant parameters are $\omega_0=0.0818$,
$\gamma=8.01 \cdot 10^{-4}$,  $\Omega=0.082$,
$\Gamma=4.58 \cdot 10^{-4}$.}
\end{figure}

The simplest  choice, in our opinion, is to take
\begin{equation}\label{eq:funzione_prova}
\langle P(0)\dot P(t)\rangle \propto-  \frac {\sin \omega_0 t}{\cosh \gamma
  t} \ ,
\end{equation}
which satisfies all the requirements. It is then an easy task to compute the
imaginary part of $\hat \chi(\omega)$ through the residue theorem, and one finds
\begin{equation}
\label{eq:2n}
\mathrm{Im}\, \hat \chi(\omega) \propto\frac {1}{\gamma} \left(
{\sech\left(\frac{\pi(\omega-\omega_0)
  }{2\gamma}\right)}-
{\sech\left(\frac{\pi(\omega+\omega_0)}{2\gamma}\right
  )} \right) \ ,
\end{equation}
As can be easily checked, this expression gives an exponential decay at
high frequencies, while reducing to a Lorentzian for $\omega$ near
$\omega_0$ .\footnote{One has to recall that $\sech x \Def \frac
1{\cosh x} \simeq    (1+x^2/2)^{-1}$ for $x\simeq0$. Obviously we are
in the case in which the line 
  width is much smaller than the frequency of the peak, so that at
  most one of the terms is non-negligible in expression (\ref{eq:2n}). }
\begin{figure}[!t]
~\hskip -1.7truecm\begin{egp}
  set ylabel "Real part of ~{/Symbol c}{.1\303} "
  set xlabel "Wavenumber [cm^{-1}]"
  set ytics -6,2,6
  plot [0:2500][-5:5]"uscita_LiF.dat" i 0 u 1:(($2+chiLi)/(4*pi)) w l not,\
  "LiF_dati.txt" u 2:(($4>0 && $5>0) ? (($4*$4-$5*$5-1)/(4*pi)) : 1/0)\
  w p pt 6   not  #$
\end{egp}
\begin{egp}
set log y
set ylabel "Imaginary part of ~{/Symbol c}{.1\303} "
set xlabel "Wavenumber [cm^{-1}]"
set ytics 1e-8,100,100
plot [0:2500] [1.E-8:100] "uscita_LiF.dat" i 0 u 1:($3/(4*pi)) w l not,\
"LiF_dati.txt" u 2:(($4>0 && $5>0) ? (2*$4*$5/(4*pi)) : 1/0) w p pt 6\
not #$
\end{egp}
\caption{ Real (left) and imaginary (right) parts of  susceptibility
  versus frequency for a \ce{LiF}  crystal. Circles are experimental
  data taken from \cite{palik}. Solid lines represent real and
  imaginary parts
  of the fitting function  (\ref{eq:fit}), with constants chosen as in
  Table~\ref{tab:1} and $\chi_\infty=7.64\cdot 10^{-2}$, as given by \cite{kachare}.\label{fig:LiF}}
\end{figure}

It is instead impossible to find a close expression for the real part
of $\hat \chi(\omega)$. However some approximating expansions can be
found, starting from the expansions (see \cite{skh}, p.
191) for the Laplace transform of the hyperbolic secant
\begin{equation}
\label{eq:espansione}
\mathscr{L}(s) \Def \int_0^{+\infty} \frac {e^{-st}}{\cosh t} \diff t = 2
\sum_{k=0}^{\infty} \frac {(-1)^k}{s + 2k + 1} =
\cfrac{1}{s+\cfrac{1}{s+ \cfrac{4}{s+\cfrac{9}{\cdots}}}}\ .
\end{equation}
The first expansion at the r.h.s. should be used for small $s$, i.e.,
near the peak, 
while for larger values of $s$ the continued fraction expansion should
be used. On the other hand one can check by inspecting
Figure~\ref{fig:2}, that a good approximation for the real part of
susceptibility is given by  
\begin{equation}
\label{eq:1n}
\textrm{Re}\, \hat \chi(\omega) \simeq \Omega A'\, 
\frac {\Omega^2 - \omega^2}{(\Omega^2 -  \omega^2)^2 + 4\Gamma^2\omega^2}
\end{equation}
with suitably chosen constants $A'$, $\Omega$ and $\Gamma$. In particular
$\Omega$ is very close to $\omega_0$, while $\Gamma$ turns out to be
smaller then $\gamma$. 

So we propose that formula (\ref{eq:2n}) should be used in place of
(\ref{eq:1im}) in fitting the experimental data for the imaginary part
of $\hat \chi$, which are those actually
obtained from the experimental values of $n$ and $\kappa$. As an
example, we have selected three relevant cases of ionic crystals, and 
show below that good fits are obtained. Moreover, the
parameters entering the fit provide a good approximation also for
the real part of susceptibility, as expected.

\begin{figure}[!t]
~\hskip -1.7truecm\begin{egp}
  set ylabel "Real part of ~{/Symbol c}{.1\303} "
  set xlabel "Wavenumber [cm^{-1}]"
  set ytics -6,2,6
  plot [0:1000][-5:5] "uscita_NaCl.dat" i 0 u 1:(($2+chiNa)/(4*pi)) w l not,\
  "NaCl_dati.txt" u 2:(($4>0 && $5>0) ? (($4*$4-$5*$5-1)/(4*pi)) : 1/0) w p pt  6 not #$
\end{egp}
%
\begin{egp}
set log y
set ylabel "Imaginary part of ~{/Symbol c}{.1\303} "
set xlabel "Wavenumber [cm^{-1}]"
set ytics 1.e-8,100,100
plot [0:1000] [1.E-8:100] "uscita_NaCl.dat" i 0 u 1:($3/(4*pi)) w l not,\
"NaCl_dati.txt" u 2:(($4>0 && $5>0) ? ((2*$4*$5)/(4*pi)) : 1/0) w p pt\
6  not #$
\end{egp}
\caption{Real (left) and imaginary (right) parts of susceptibility
  versus frequency for a \ce{NaCl}  crystal. Circles are experimental
  data taken from \cite{palik}. Solid lines represent real and
  imaginary parts
  of the fitting function  (\ref{eq:fit}), with constants chosen as in
  Table~\ref{tab:1} and $\hat\chi_\infty=1.08\cdot 10^{-1}$.\label{fig:NaCl}}
\end{figure}

\section{Fit of the susceptibilities of selected
crystals}\label{sez:fit}

\begin{table}[!t]
  \begin{center}
    \begin{tabular}{|l||c|c|c|c|c|c|c|}
      \hline
      \multicolumn{8}{|c|}{\ce{LiF}} \\
      \hline
      $A_k$      & 137 & 27.8  &  9.36 &  3.7 & 2.29 &  \quad&\qquad \\ 
      \hline       
      $\omega_k$ & 307 & 307  & 307  & 500  & 110  &  \quad&\qquad \\
      \hline
      $\gamma_k$ & 12.6 & 55 &  220 & 59 & 50  &  \quad&\qquad \\
      \hline
      \multicolumn{8}{c}{~} \\
      \hline
      \multicolumn{8}{|c|}{\ce{NaCl}} \\
      \hline
      $A_k$      & 39.8 & 6.37 & 5.57 & 0.61 & 0.56 & 0.193 &\\ 
      \hline
      $\omega_k$ & 164 & 155 & 161 & 254 & 233  & 296 &\\
      \hline
      $\gamma_k$ & 3.9 & 79 & 11  & 11 & 11  &  17 &\\
      \hline      
   \multicolumn{8}{c}{~} \\
      \hline
      \multicolumn{8}{|c|}{\ce{MgF2} } \\
      \hline
      $A_k$      &40.1& 38.2 & 7.32 & 5.41 & 3.63 & 3.06 & 1.59\\ 
      \hline
      $\omega_k$ & 248 &450  &407  &435  &420   &485 &243  \\
      \hline
      $\gamma_k$ & 6.3 & 9.4 & 6.3  & 27 & 188  &  71 & 39 \\
      \hline
    \end{tabular}
  \end{center}
  \caption{\label{tab:1} Fitting constants for the function
    (\ref{eq:fit}), for the three selected substances. }
\end{table}

We present here,  for three selected elements, the fits of the real
and the imaginary parts of
susceptibility for three much studied ionic crystals, i.e., \ce{LiF},
\ce{NaCl} and \ce{MgF2}.  As, in general, several lines could be present,
we choose to fit the data with the following function
\begin{equation}\label{eq:fit}
  \begin{split}
     \hat \chi(\omega) &= \hat\chi_{\infty} + \sum_{k=1}^N \frac{A_k}{2i\gamma_k} \left[ 
      \mathscr{L}\left(i \frac {\omega+\omega_k}{\gamma_k} \right) 
    - \mathscr{L}\left( i\frac {\omega-\omega_k}{\gamma_k}\right) 
     \right]       \ ,
  \end{split}
\end{equation}
where $i$ is the imaginary unit,  $N$ is the number of terms that
should be chosen in order to match the 
experimental data, $\mathscr{L}(\cdot)$ is the function defined
in (\ref{eq:espansione}), while $\hat\chi_{\infty}$ is the electronic
contribution 
to susceptibility which, in the infrared region, just reduces to a real
constant.  The values of the parameters we found are summarized
in Table~\ref{tab:1}, while in Figures~\ref{fig:LiF}--\ref{fig:MgF2} 
we plot both the experimental data and the curves found. 

For what
concerns the experimental data, we recall that only  the values of
$n$ and $\kappa$,  and not those of the complex susceptibility $\hat\chi$,
are usually reported in the literature (see \cite{palik} 
and the references therein). Thus, the complex susceptibility 
has to be recovered  from the tabulated values of $n$ and
$\kappa$: this is simple if, for a given frequency, the values of both $n$
and $\kappa$ are tabulated (see footnote~\ref{foot:1}). As this is
not always the case in the region where $n$ is almost constant,
we estimated the refractive index $n$
by linear interpolation when needed.
\begin{figure}[!t]
~\hskip -1.7truecm\begin{egp} 
  set ylabel "Real part of ~{/Symbol c}{.1\303}"
  set xlabel "Wavenumber [cm^{-1}]"
  set ytics -4,2,4
  plot [0:2100][-4:4] "uscita_MgF2.dat" i 0 u 1:(($2+chiMg)/(4*pi)) w l not,\
  "MgF2_dati.txt" u 2:(($4>0 && $5>0) ? (($4*$4-$5*$5-1)/(4*pi)) : 1/0) w p pt 6 not #
\end{egp}
\begin{egp}
set log y
set ylabel "Imaginary part of ~{/Symbol c}{.1\303}"
set xlabel "Wavenumber [cm^{-1}]"
set ytics 1e-8,100,100
plot [0:2100] [1.E-8:100] "uscita_MgF2.dat" i 0 u 1:($3/(4*pi)) w l not,\
"MgF2_dati.txt" u 2:(($4>0 && $5>0) ? ((2*$4*$5)/(4*pi)) : 1/0) w p\
pt 6 not #$
\end{egp}
\caption{Real (left) and imaginary (right) parts of susceptibility
  versus frequency for the ordinary ray in a \ce{MgF2}
  crystal. Circles are experimental 
  data taken from \cite{palik}. Solid lines represent real and
  imaginary parts
  of the fitting function  (\ref{eq:fit}), with constants chosen as in
  Table~\ref{tab:1} and $\hat\chi_\infty=7.08\cdot 10^{-2}$.\label{fig:MgF2}}
\end{figure}
\end{egpfile}

Two remarks are in order: the first one is that our procedure to determine
the parameters, which involves only the
imaginary part of susceptibility given by expression
(\ref{eq:2n}),  was just that of trial and error. In other terms,  we find by hands
some values of the parameters which, in our opinion, give acceptable
fits for the experimental data over a large range of values of $\hat
\chi$ (nine orders of magnitude). \textbf{No} procedure of error
minimization, nor any statistical test, are used to check the quality of
the fit. This in 
particular implies that also the number of terms $N$ in the sum is taken
in a sense in an arbitrary way. This point will be discussed below.
The data of the real part of susceptibility, which do not enter the
fit, are then used for determining the constant 
$\hat\chi_{\infty}$. 

The second remark is that most of the ``experimental data'' in the region of
the ``peaks'', are not experimental at all, because in such a region neither
the extinction coefficient nor the refractive index can be actually measured.
They are actually inferred from other measured quantities (such as the
reflectivity) assuming that susceptibility can be described in a fairly
good way by the Lorentz model. This is particularly evident in
Figure~\ref{fig:LiF}, in which two sets of data are present in the
region between 800 cm$^{-1}$ and  1000 cm$^{-1}$. The set of values which
presents an exponential decay corresponds to the directly measured
values of $\kappa$, while the other one contains the values
which are inferred from reflectivity measurements.

In view of these two remarks, we expect that the overall accuracy of the fits
would
be improved is a reprocessing were performed of the experimental data,
in the spectral region where the extinction coefficient and the refractive
index are not directly measurable.


\section{A final Comment}\label{sez:com}
We discuss here the problem of the number $N$ of
terms to be used in the expression (\ref{eq:fit}) for the fit of the
imaginary part of susceptibility. Naively, one might expect $N$
to be equal to the number of most ``evident'' lines, but actually a larger
number of terms is needed: in particular
(as shown by Table~\ref{tab:1}) five terms are needed to match the data for \ce{LiF},
six for \ce{NaCl} and seven for \ce{MgF2}. For example, in order to
have a good global fit three different terms have to be associated to the
 \ce{LiF} line at $307$ cm$^{-1}$, all with the same frequency, but with
three well different damping constants. Something analogous  occurs for the
\ce{NaCl} line at about 160 cm$^{-1}$  and for the \ce{MgF2} line at about
245 cm$^{-1}$, for which several terms of very near frequencies are
needed, with however different values of $\gamma$.

This can be interpreted by saying that several time--scales are
involved in the decay of the relevant correlations, so that the
relaxation of the correlations is a much more complicated process than
just a simple exponential decay. 

This is a well known fact in other fields, for example in glasses
(see \cite{vetri}), or in relaxation spectroscopy (see for example
\cite{dissado}), which involve the 
behavior of susceptibility at low frequencies (in the micro-waves
region), where continuous distributions of relaxation
times are actually used. So our result could be read as a hint that,
also in the infrared 
region, the process of the decay of correlations is a complicated one. And
indeed, the numerical evidence in some cases seems to support
this view. We refer to computations performed by us in the case of a one
component model of plasma, see \cite{plasmi}, and of a model
of \ce{LiF} crystal, see \cite{gangemi}.\footnote{In the cited paper we
  do not report explicitly the correlation as a function of time, but
  an example of such a plot can be found at the following URL: 
  \texttt{http://fgangemi.unibs.it/LiF\_results/avg-corr\_na4096-T300\_10sim.png}  }
In any case, more work is needed to set this point. 

\vskip 1.em
\noindent
\textbf{Acknowledgments}. We thank prof. L.~Galgani for  thorough
discussions and for his invaluable help in correcting the manuscript.

\end{document}